\documentclass[showpacs, twocolumn]{revtex4-1}
\usepackage{graphicx}
\usepackage{amsmath}
\usepackage{csquotes}

\usepackage[usenames,divipsnames,svgnames,table]{xcolor}
\usepackage[unicode=true,pdfusetitle,
bookmarks=true,bookmarksnumbered=false,bookmarksopen=false,
breaklinks=true,pdfborder={0 0 0},backref=false,colorlinks=true]{hyperref}
\hypersetup{linkcolor=NavyBlue,urlcolor=NavyBlue,citecolor=NavyBlue}
\usepackage{breakurl}

\begin{document}

\title{Quantum information flow in impurity qubits interacting with Bose-Bose mixtures}

\author{ Abdel\^{a}ali Boudjem\^{a}a$^{1,2}$}
\email {a.boudjemaa@univ-chlef.dz}
\author{Lan Xu$^3$}
\author{Qing-Shou Tan$^{4}$}
\email{qstan@hnist.edu.cn}
\affiliation{$^1$Department of Physics, Faculty of Exact Sciences and Informatics, 
and $^2$Laboratory of Mechanics and Energy, Hassiba Benbouali University of 
Chlef, P.O. Box 78, 02000, Chlef, Algeria.\\
$^3$School of Physics and Chemistry, Hunan First Normal University, Changsha 410205, China.\\ 
 $^4$Key Laboratory of Hunan Province on Information Photonics and 
Freespace Optical Communication,
College of Physics and Electronics, Hunan Institute of Science and Technology, Yueyang 414000, China}

\date{\today}

\date{\today}

\begin{abstract}

We investigate the dynamics of quantum information flow in one and two impurity qubits trapped in a double-well potential and interacting with a one-dimensional ultracold Bose-Bose mixture reservoir. For a single qubit immersed in a binary Bose mixture, we show that the system maintains coherence over finite timescales and exhibits non-Markovian dynamics, particularly in the upper branch of the environment. We explore the transition from Markovian to non-Markovian dephasing through the Ohmicity of the spectral density functions, which are significantly influenced by interspecies interactions.
In the case of two spatially separated qubits coupled to the Bose-Bose mixture reservoir, we demonstrate that collective decoherence affects the system dynamics, leading to prolonged coherence survival in both branches of the mixture. The intricate evolution of the decoherence factors is reflected in the density spectral functions and their Ohmicity characteristics. We find that the decoherence functions and spectra oscillate with increasing distance between the qubits, modifying the information flow dynamics. Additionally, we conduct a thorough investigation of the entanglement dynamics between the two qubits induced by the binary Bose mixture reservoir in both branches, underscoring the critical role of interspecies interactions.

\end{abstract}


\maketitle

\section{Introduction} 

In recent years, quantum information has attracted much attention due to its potential applications in quantum computing, communication, metrology and simulation.
With their unprecedented degree of controllability, their low temperatures, and their ability to create entangled states of many particles, trapped ultracold atoms constitute 
ideal platforms to investigate quantum information.

Trapped neutral-atom qubits are among the most promising candidates for qubit encoding in quantum information processing due to their 
identical characteristics and long coherence times. 
Recently, mixed systems composed of impurity qubits trapped in a double-well potential or with an internal level structure trapped in a deep harmonic potential
interacting with a reservoir consisting of Bose-Einstein condensates (BECs) have attracted a great deal of interest both theoretically and experimentally  (see e.g. \cite{Kang,Reca,Ciro,Breu,Haik,Bru1,Will,Bru2,Speth, Haik1,Tang1,Hang,Jev,Hauke, Sev,Kil,Song,Fey,Mah,Mit,Mis,Khan,Boudj8}).
Furthermore, the collective dephasing of two impurity qubits immersed in a single BEC reservoir was studied in \cite{Ciro}.
The non-Markovianity and the entanglement of such two spatial states impurity were addressed in (see e.g. \cite{Xu, Addis, McE, Tan2} and references therein).

Meanwhile, the ability to control interactions using Feshbach resonances has led to the creation of two-component BECs in two different spin states (see e.g \cite{Mya, Hall, Stam,Sten,Err1,Marte}.
One of the most elusive features of binary BECs is their miscibility-immiscibility transition which depends on  
the ratio of the intra- and interspecies interactions providing an understanding of the interplay between the two Bose components.
Bose mixtures have been used to probe many aspects including topological defects \cite{Kev,Matt}, 
spin-orbit coupling \cite{Chen},  quantum and thermal fluctuations \cite{Roy, Boudj1, Ota1}, localization in disordered two-component BECs \cite{Boudj4,Boudj5,Abbas}, and droplet states  
(see e.g \cite{Petrov,Cab, Sem, Err, Boudj6} and references therein).
Binary BECs have  also proven to be a fruitful testbed for storing and manipulating quantum information (see e.g \cite{Li, Byr} and references therein).

Recent advances in the field of cold atoms have sparked interest in Bose-Bose-impurity mixtures \cite{Compa, Boudj7,Bighin}.
The coherent exchange between the constituents of such exotic systems allows for the necessary high level of control and provides entangled qubit states offering an ideal platform
for quantum information processing. 
Our aim in this paper is then to propose a scheme to enhance the control of quantum information flow in impurity qubits interacting with ultracold Bose-Bose mixtures.
The presence of the dual BECs surrounding the impurities  leads us to tune the induced environmental noise, giving rise to non-dissipative decoherence due to collisions 
between atoms of the two species and the impurity atoms.
The main difference between such a two-component BEC reservoir, which accommodates two branches of excitations, namely,
the upper branch (which has a density nature) and the lower branch (which has a spin nature), and other proposals for quantum information processing is the manner and the nature of
how qubit information is encoded in each component.

First, we investigate a system consisting of a single impurity qubit trapped in a double-well potential immersed in a  quasi-one-dimensional (quasi-1D) two-component BEC 
in two different hyperfine states. One of the advantages of such a double-well potential is that it is characterized by a separation between the two minima of the impurity double well,
which may lead to lower decoherence due to low-frequency modes of the environment.
We calculate the decoherence factors, and the non-Markovianity measure using  the exact pure dephasing  model.
Furthermore, we also discuss the link between the occurrence of memory effects and the form of the spectral density function in each component.
We show that the qubit coupled to both the upper and lower branches retains its coherence at long times inducing a non-Markovian dynamics,
although the coherence in the upper branch is more robust.
We find that the spectral density  functions exhibit rich structures in terms of the interspecies interaction strength
allowing transitions from sub-Ohmic to Ohmic to super-Ohmic in the  low-frequency regime. 
Therefore,  a crossover from Markovian to non-Markovian dynamics takes place in a double-well single qubit.
For large separation between wells, the spectra oscillate for any interspecies interaction strength with different amplitudes in different branches.

In addition, we address another interesting question of quantum information with open quantum systems, namely, 
the effects of collective decoherence of a system of two spatially separated qubits and dephasing under the influence of a two-component BEC in a quasi-1D configuration.
We specifically examine how qubit-qubit correlations induced by the Bose mixture reservoir influence the degree of non-Markovianity, which is quantified based on the negativity of the decay rates.
Our numerical results reveal the existence of super- and subdecoherent two-qubit states in both branches of the mixture, similar to those found in two-qubit interactions with 
a single-component BEC environment \cite{Ciro}.
Importantly, we observe an appealing dynamical crossover from Markovian nature to non-Markovian behavior in the lower branch when the interspecies interaction is increased.
Moreover,  we find that the interplay of interspecies interaction and the correlation between qubits may lead to unusual and interesting phenomena, 
like oscillations of coherence at finite times and the survival of coherence at long times in both lower and upper branches.
We show that changing the strength of the interspecies interaction leads to nontrivial spectral density functions, altering the Ohmicity character of the system.
The relevance of the spatial separation between qubits in  the dynamics of the system is also highlighted.
For large separation between wells, the density spectral functions experience oscillations with high and small amplitudes alternately. 

Finally, we investigate the entanglement dynamics of two coupled qubits, starting from a non-entangled state induced by the BEC reservoir, which is a crucial aspect of quantum information. 
By solving the system's dynamics, we analyze the time evolution of concurrence, revealing how environment-induced interactions affect the entanglement between the qubits. 
We emphasize that interspecies interactions can enhance entanglement in one component while reducing it in the other over a short time interval.

The rest of this paper is organized as follows. Section \ref{model} introduces the model of impurity qubits immersed in a quasi-1D Bose-Bose mixture reservoir. 
We present the interaction between the impurity qubits and the two distinct branches of the reservoir, along with a representation of the decoherence function. 
Section \ref{singq} is dedicated to analyzing the information flow in the case of a single qubit by calculating the non-Markovian effects. 
In Secs.~\ref{twoq} and \ref{Entg}, we further explore the effects of collective decoherence in the case of two qubits and the entanglement dynamics induced by the reservoir. 
Finally, we present our conclusions and discussions in Sec.~\ref{Conc}.

\section{Model} \label{model}

\begin{figure}
\includegraphics[scale=0.6] {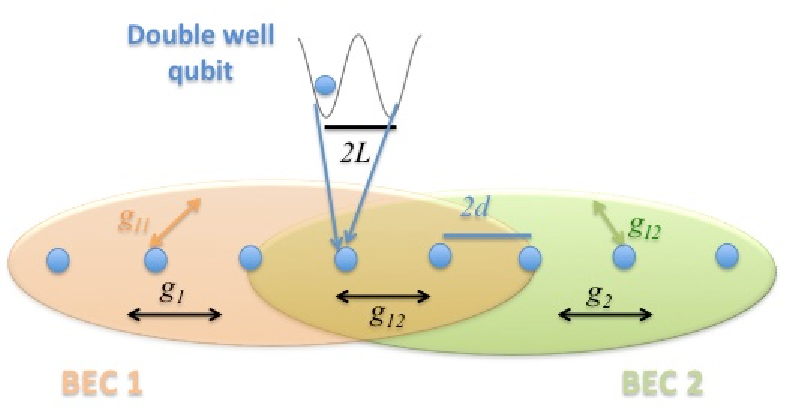}
 \caption{ A mixture of two BECs confined in a very shallow potential interacts with  impurity atomic qubits, each of which is trapped in a double-well potential.
The distance between two wells in the same trap is $2L$ (i.e. the size of the qubit) and the distance between adjacent traps is $2d$.}
 \label{schm}
\end{figure}

We consider a few impurity atoms of mass  $m_I$  embedded in a a thermally equilibrated two-component BEC  with atomic mass $m_j$ in a quasi-1D harmonic confinement 
for repulsive impurity-BEC couplings at temperature $T$. 
The atomic qubit is trapped in a double-well potential and forms an effective qubit system in which the two
qubit states are represented by occupation of the impurity in the left $|l \rangle$ or the right $|r \rangle$ well of double well $i$.
The distance between two wells in the same trap is $2L$ and the distance between adjacent traps is $2d$ \cite{Ciro}  (see Fig.~\ref {schm}).
The model was introduced first in Refs. \cite{Palma, Rein, Ciro} for two-qubit dephasing under the influence of a single BEC reservoir.

The Hamiltonian of the qubits can be written in terms of the Pauli operators, $\hat \sigma_z=| l \rangle \langle l|-| r \rangle \langle r|$, as:
\begin{equation} \label{ham11}
\hat H_{\text I}= \frac{\hbar}{2} \sum_i\left[\Omega _r^i \left(1+\hat \sigma_z^i\right)+\Omega _l^i \left(1-\hat \sigma_z^i\right)\right],
\end{equation}
where $\hbar\Omega _{l,r}^i$ are the energies of the impurity atoms in the double wells $i$.

For the two-component BECs (reservoir) with mass $m_\nu$, where the index $\nu=1,2$ refers to the species label, we assume that they are harmonically confined in a quasi-1D geometry. 
In this case, the scattering lengths characterizing the intra- and interspecies interactions are obtained from the three-dimensional ones via
$g_\nu =2 \hbar^2 a/(m_{\nu} l_{\nu}^2)$, and $g_{12}= 2\hbar^2 a_{12}/ [m_{12} (l_1^2+ l_2^2)]$, respectively,
where  $a_{\nu}$ and $a_{12}$ are intra- and interspecies $s$-wave scattering lengths, $m_{12}= m_1 m_2/(m_1+m_2)$ is the bosonic reduced mass,
$l_{\nu}= \sqrt{\hbar/m_{\nu}\omega_{\nu \perp}}$, and $\omega_{\nu \perp}$ is the transverse trapping frequency which should be much larger than the longitudinal  trapping frequency 
$\omega_{\nu x}$ i.e. $\omega_{\nu x}/\omega_{\nu\perp}\ll1$.  
Here the impurities are supposed to be pinned in the center of the trap.  
The Bogoliubov theory suggests that the Hamiltonian of the reservoir must be written as:
\begin{equation}  \label{ham2}
\hat H_{\text{BB}}\!\!= \sum_{\mathbf k \neq 0} \left[\varepsilon_{+ \mathbf k} \hat b^\dagger_{1\bf k} \hat b_{1\bf k} + \varepsilon_{- \mathbf k} \hat b^\dagger_{2\bf k} \hat b_{2\bf k}\right],
\end{equation}
where $\hat b^\dagger_{\nu \bf k}$ and $\hat b_{\nu \bf k}$ are operators of elementary excitations obeying the usual Bose commutation relations, and 
the Bogoliubov excitations read
\begin{equation} \label{Bog}
\varepsilon_{\pm k}= \sqrt{ \frac{\varepsilon_{1k}^2 +\varepsilon_{2k}^2} {2}  \pm \sqrt{ \left(\frac{\varepsilon_{1k}^2 -\varepsilon_{2k}^2} {2}\right)^2 + 4 g_{12}^2 n_1 n_2 E_{1k} E_{2 k}  } },
\end{equation}
where $\varepsilon_{\nu k}= \sqrt{ E_{\nu k}^2 +2E_{\nu k} g_{\nu} n_{\nu}}$ are the Bogoliubov spectra for the individual components,
and $E_{\nu k}=\hbar^2k^2/(2m_{\nu})$ is the energy of the free particle in each component.
The Bogoliubov spectrum  (\ref{Bog}) is composed of two branches namely: the highest-energy branch $\varepsilon_{k+}$  and lowest-energy branch $\varepsilon_{k-}$.
Both branches exhibit a phononlike linear dependence on $k$ in the long-wavelength limit 
\begin{equation} \label{sound}
\varepsilon_{k\pm}= \hbar c_{s\pm} k, 
\end{equation}
where
$$c_{s\pm}^2= \frac{1}{2}  \bigg(c_{s1}^2+c_{s2}^2 \pm \sqrt{ \left(c_{s1}^2-c_{s2}^2 \right)^2+ 4 c_{s12}^2}\bigg),$$
are the sound velocities of the upper and lower branches, with 
$c_{s \nu} = \sqrt{g_{\nu} n_{\nu }/m_{\nu}}$  being the sound velocities for each species and $c_{s12}= \sqrt{g_{12} ^2n_1 n_2/m_1 m_2}$.
The stability condition requires: $ c_{s1} c_{s2} >  c_{s12}^2$.

The interaction between the qubits and the binary BECs reservoir is described by a contact pseudopotential,
$V_{I\nu} (x)= g_{I \nu}\delta (x)$, where $g_{I \nu}= 2\hbar^2 a_{I \nu}/ m_{I\nu} (l_I^2+ l_{\nu}^2) $, and $m_{I\nu}= m_{\nu}m_I/(m_{\nu}+m_I)$ is the impurity-host reduced mass.
Therefore, the qubit-reservoir coupling Hamiltonian is defined  as:
\begin{align} \label{ham3}
\hat H_{\text{IBB}} &= \sum_{\mathbf k \neq 0} \sum_i \bigg[ \left( g_{+\mathbf k}^{ir} \hat b_{1\bf k}+g_{+\mathbf k}^{i r*}\hat b^\dagger_{1\bf k}\right) (1+\hat \sigma_z^i) \nonumber\\
&+ \left(g_{+ \mathbf k}^{i l} \hat b_{1\bf k}+\tilde g_{+\mathbf k}^{i l*}\hat b^\dagger_{1\bf k}\right) (1-\hat \sigma_z^i) \bigg] \nonumber\\
&+\sum_{\mathbf k \neq 0} \sum_i \bigg[ \left( g_{-\mathbf k}^{ir} \hat b_{2\bf k}+ g_{-\mathbf k}^{i r*}\hat b^\dagger_{2\bf k}\right) (1+\hat \sigma_z^i) \nonumber\\
&+ \left( g_{- \mathbf k}^{i l} \hat b_{2\bf k}+ g_{-\mathbf k}^{i l*}\hat b^\dagger_{2\bf k}\right) (1-\hat \sigma_z^i) \bigg],  
\end{align}
where the qubit-reservoir coupling parameter in each component reads
\begin{subequations} \label{srcp}
\begin{align}  
g_{+ \mathbf k}^{i(r,l)}=g_{I1} \sqrt{\frac{n_1}{\ell}} \sqrt{\frac{E_{1k}}{\varepsilon_{+ \mathbf k}} } \int d {\mathbf x } |\phi^i ({\mathbf x}_{r,l})|^2 e^{i {\mathbf k}\cdot {\mathbf x}},\\
g_{- \mathbf k}^{i(r,l)}=g_{I2} \sqrt{\frac{n_2}{\ell}} \sqrt{\frac{E_{2k}}{\varepsilon_{- \mathbf k}} } \int d {\mathbf x } |\phi^i ({\mathbf x}_{r,l})|^2 e^{i {\mathbf k}\cdot {\mathbf x}},
\end{align} 
\end{subequations}
where
$\phi^i$ are the real eigenstates of impurity atoms localized in  the two wells of the potential, and 
$\ell$ is the size of the quasi-1D condensate.

The total Hamiltonian of the system is given by the combination of three sub-Hamiltonians: $\hat{H} = \hat{H}_{\text{I}} + \hat{H}_{\text{BB}} + \hat{H}_{\text{IBB}}$. 
According to Eqs. (\ref{ham3}) and (\ref{srcp}), the coupling strength between the qubit and each component is related to the $s$-wave scattering strength, which can 
be precisely controlled using Feshbach resonance technology. This allows us to effectively select the component with which the qubit couples to the BEC reservoir. 
Consequently, we will next discuss the impact of each BEC reservoir component on the qubit's information flow dynamics.

The evolution operators $\hat{U}_{\pm}(t) = \exp(-i \hat{H}_{\pm} t)$, which are generated by the Hamiltonian of the entire system for each branch $\hat{H}_{\pm}$, are expressed as follows:
\begin{widetext}
\begin{subequations}\label{EO}
\begin{align} \label{EO+}
\hat U_{+} (t)&= \exp { \bigg[ -i  \sum_{\mathbf k } \varepsilon_{+ \mathbf k } \hat b_{1\mathbf k} ^\dagger  \hat b_{1\mathbf k} \,t \bigg]} 
\exp \bigg[\sum_{\mathbf k } \left( \sum_{i} A_{+ \mathbf k } ^i (t) \hat \sigma_z^i + \beta_{+ \mathbf k }  (t) \right) \hat b^\dagger_{1\mathbf k } 
-\sum_{\mathbf k } \left( \sum_{i}  A_{+ \mathbf k } ^{i*} (t) \hat \sigma_z^i + \beta^*_{+ \mathbf k } (t)\right) \hat b_{1\mathbf k} \bigg] \nonumber\\
&\times \exp{\bigg[ i \sum_{\mathbf k} f_{+ \mathbf k } (t) \Re \sum_{ij} \frac{(g_{+ \mathbf k}^{i r} -g_{+ \mathbf k}^{i l}) (g_{+ \mathbf k}^{jr *}-g_{+ \mathbf k}^{jl*}) }
{ 4 \varepsilon_{+ \mathbf k} ^2} \hat \sigma_z^i \hat \sigma_z^j \bigg]} 
 \exp{\bigg[ i \sum_{\mathbf k} f_{+ \mathbf k } (t) \Re \sum_{i} \frac{(g_{+ \mathbf k}^{ir} -g_{+ \mathbf k}^{i l} ) \sum_{j}  (g_{+ \mathbf k}^{j r*}+g_{+ \mathbf k}^{jl*}) }
{ 2 \varepsilon_{+ \mathbf k } ^2} \hat \sigma_z^i \bigg]} \nonumber\\
&\times \exp{\bigg[ i \sum_{\mathbf k} f_{+\mathbf k} (t)  \sum_{i} \frac{(g_{+ \mathbf k}^{i r} +g_{+ \mathbf k}^{i l} ) \sum_{j}  (g_{+ \mathbf k}^{jr*} +g_{+ \mathbf k}^{jl*}) }
{ 4 \varepsilon_{+ \mathbf k } ^2}\bigg]}, 
\end{align}
\begin{align} \label{EO-}
\hat U_{-} (t)&= \exp { \bigg[ -i  \sum_{\mathbf k } \varepsilon_{- \mathbf k } \hat b_{2\mathbf k} ^\dagger  \hat b_{2\mathbf k} \,t \bigg]} 
\exp \bigg[\sum_{\mathbf k } \left( \sum_{i} A_{- \mathbf k } ^i (t) \hat \sigma_z^i + \beta_{- \mathbf k }  (t) \right) \hat b^\dagger_{2\mathbf k } 
-\sum_{\mathbf k } \left( \sum_{i}  A_{- \mathbf k } ^{i*} (t) \hat \sigma_z^i + \beta^*_{- \mathbf k } (t)\right) \hat b_{2\mathbf k} \bigg] \nonumber\\
&\times \exp{\bigg[ i \sum_{\mathbf k} f_{- \mathbf k } (t) \Re \sum_{ij} \frac{(g_{- \mathbf k}^{i r} -g_{- \mathbf k}^{i l}) (g_{- \mathbf k}^{jr *}-g_{- \mathbf k}^{jl*}) }
{ 4 \varepsilon_{- \mathbf k} ^2} \hat \sigma_z^i \hat \sigma_z^j \bigg]} 
 \exp{\bigg[ i \sum_{\mathbf k} f_{- \mathbf k } (t) \Re \sum_{i} \frac{(g_{- \mathbf k}^{ir} -g_{- \mathbf k}^{i l} ) \sum_{j}  (g_{- \mathbf k}^{j r*}+g_{- \mathbf k}^{jl*}) }
{ 2 \varepsilon_{- \mathbf k } ^2} \hat \sigma_z^i \bigg]} \nonumber\\
&\times \exp{\bigg[ i \sum_{\mathbf k} f_{-\mathbf k} (t)  \sum_{i} \frac{(g_{- \mathbf k}^{i r} +g_{- \mathbf k}^{i l} ) \sum_{j}  (g_{- \mathbf k}^{jr*} +g_{- \mathbf k}^{jl*}) }
{ 4 \varepsilon_{- \mathbf k } ^2}\bigg]}, 
\end{align}
\end{subequations}
where $f_{\pm \mathbf k } (t)=\varepsilon_{\pm \mathbf k} t -\sin{(\varepsilon_{\pm \mathbf k } t})$, 
$A_{\pm \mathbf k } = \left(1-e^{i \varepsilon_{\pm \mathbf k } t}\right) (g_{\pm \mathbf k}^{ir*} -g_{\pm \mathbf k}^{i l*})/ (2\varepsilon_{\pm \mathbf k}) $, and
$\beta_{\pm \mathbf k} = \left(1-e^{i \varepsilon_{\pm \mathbf k } t}\right) (g_{\pm \mathbf k}^{i r*} +g_{\pm \mathbf k}^{il*})/ (2\varepsilon_{\pm \mathbf k}) $.
\end{widetext}
In the evolution operators (\ref{EO}), the first term on the r.h.s. mainly causes the decoherence effect, while the second term induces interactions between the impurity qubits.
Obviously, for $g _{12}=0$, the operators $\hat U_{\pm} (t)$ reduce to that describing the dynamics of qubits immersed in a single BEC \cite{Ciro}.

Each qubit dephases under the effect of its binary BECs environment since $H_{\rm IBB}$ commutes with the total Hamiltonian. 
This means that the diagonal elements of the density matrix are constant while the off-diagonal elements decay
as $|\rho_{I}^{\{n_p\},\{m_p\}} (t)|= e^{-\Gamma^{\{n_i\},\{m_i\}} (t)}\rho_{I}^{\{n_p\},\{m_p\}} (0)$,  where 
 $\{n_p\} =\{ \rm n_1, n_2, n_3, \ldots\} $ represents the state of the impurities, with $n_p =0, 1$ indicating the presence of the atom in the left or right well, respectively.
To find  $\Gamma^{\{n_i\},\{m_i\}} (t) $, we assume that the density matrix of the initial state is $\rho^T_{\pm}(0) = \rho(0)\otimes \rho^B_{\pm}$, where the density matrix of the
reservoir is $\rho^B_{\pm} =\Pi_{\bf k}(1-e^{\varepsilon_{\pm \mathbf k}}/(k_B T))e^{-\varepsilon_{\pm \mathbf k} b^{\dagger}_{\bf k} b_{\bf k}}$,  with $k_B$ being the Boltzmann constant.
The decoherence exponent $\Gamma_{\pm}$ for each component is then defined as follows:
\begin{align} \label{DecF}
\Gamma_{\pm}^{\{n_i\},\{m_i\}} (t,T) &= \sum_{\mathbf k} \frac{2\sin^2 \left(\omega_{\pm \mathbf k} t/2 \right)} {\omega_{\pm \mathbf k}^2}  \coth \left(\frac{\omega_{\pm\mathbf k}}{2k_B T} \right)  \nonumber\\
&\times \left|\sum_{i} (m_i -n_i) \left(g_{\pm \mathbf k}^{i r}- g_{\pm \mathbf k}^{i l} \right) \right|^2 ,
\end{align}
where $\omega_{\pm \mathbf k}= \varepsilon_{\pm \mathbf k}/\hbar$ are the Bogoliubov frequencies. 
For $g_{12}=0$, the decoherence exponent reduces to that found for a qubit dephasing with a single BEC \cite{Ciro}.
The decoherence exponent (\ref{DecF}) encapsulates all the information concerning the time, the interaction, and temperature dependence of the decoherence process.
Note that the decoherence factor of the whole mixture is given by $\Gamma^{\{n_i\},\{m_i\}} (t,T)= \sum\limits_{\pm} \Gamma_{\pm}^{\{n_i\},\{m_i\}} (t,T)$.

Equations (\ref{DecF}) and (\ref{Bog}) show that with rising interspecies interaction $g_{12}$, $\omega_{-k}$ decreases, thus boosting the contribution of $\Gamma_-$
to the total decoherence exponent. On the other hand, $\omega_{+k}$ increases with $g_{12}$, which reduces $\Gamma_+$ (see also Fig.~\ref{SQb} (a) and (b)).

To be quantitative,  we consider from now on few impurities of ${}^{41}$K atoms in a bath of a mixture of two hyperfine states of ${}^{87}$Rb atoms ($m_1=m_2$) at zero temperature ($T=0$). 
Our simulations are performed for a Bose mixture of equal  masses, harmonic frequencies and densities (i.e. a symmetric Bose mixture). 
The parameters are set to: $n_1=n_2=n=3.6\times 10^7$ m$^{-1}$,
$a_1= a_2= a=5.3 \times 10^{-9}$ m \cite{Marte}, and  $a_I= 3.4 \times 10^{-9}$ m \cite{Err1}.
The trapping frequencies are assumed to be  $\omega_{1\perp}=\omega_{2\perp} =\omega_{\perp}=2 \pi$ kHz, then the corresponding
harmonic oscillator width is $l_1=l_2=l_0=3.4 \times 10^{-7}$ m.

\section{Single qubit}\label{singq}

Let us start by analyzing the decoherence factor and the non-Markovianity measure of a single qubit in each component.
We thus obtain
\begin{align} \label{DecFdw}
\Gamma_{\pm}^{\{0\},\{1\}} (t)\equiv \Gamma_{\pm}^0 (t)= \sum_{k} \frac{ 2g_{\pm  k}^2} {\omega_{\pm k}^2} \sin^2 \left(\frac{ \omega_{\pm k} t}{2} \right)  ,
\end{align}
where the qubit-bath coupling parameters (\ref{srcp}) turn out to be given as: 
\begin{equation} \label{coup}
 g_{\pm k}= g_{I} \sqrt{\frac{n}{\ell}} \sqrt{\frac{E_{k}}{\hbar\omega_{\pm k}} } e^{-(k l_I/2)^2} \sin(k L).
\end{equation}
In the short time limit, $\Gamma_{\pm}^0 (t)$ are proportional to $t^2$.  

For the sake of numerical simulations, we rescale energy  in terms of $\hbar \omega_{\perp}$, momentum in terms of $l_0^{-1}$,
time in terms of $\omega_{\perp}^{-1}$, and length is expressed in terms of $l_0$.
We introduce the following dimensionless parameters:  $\alpha= 4 n a$, and $p= l_I/l_0$ with $l_I=\sqrt{\hbar/m_I \omega_I}$.

We now take a closer look at the role of interspecies interactions in the decoherence factor and Non-Markovianity  measure in the upper and lower branches of the reservoir.

\begin{figure}
\includegraphics[scale=0.46] {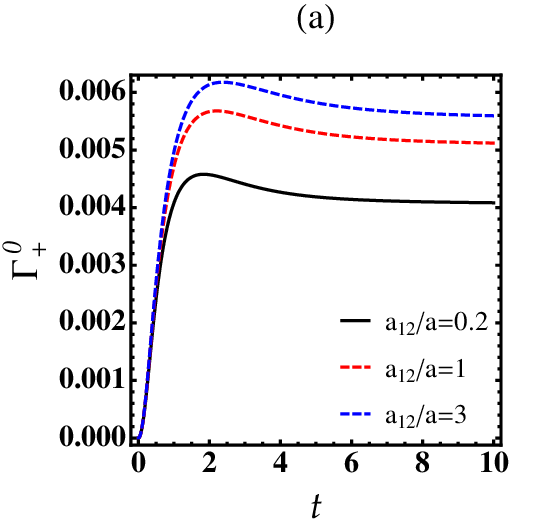}
\includegraphics[scale=0.46] {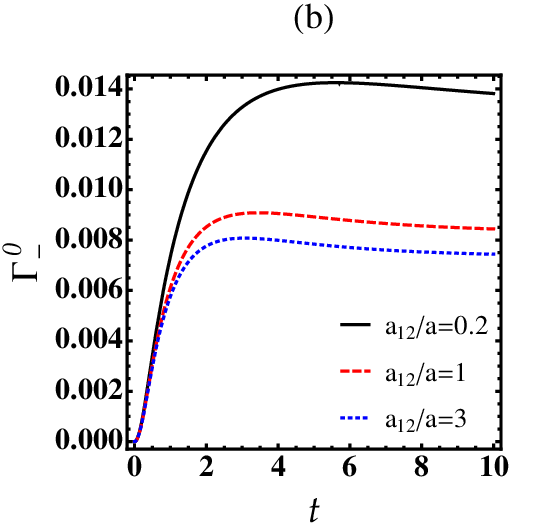}
\includegraphics[scale=0.45] {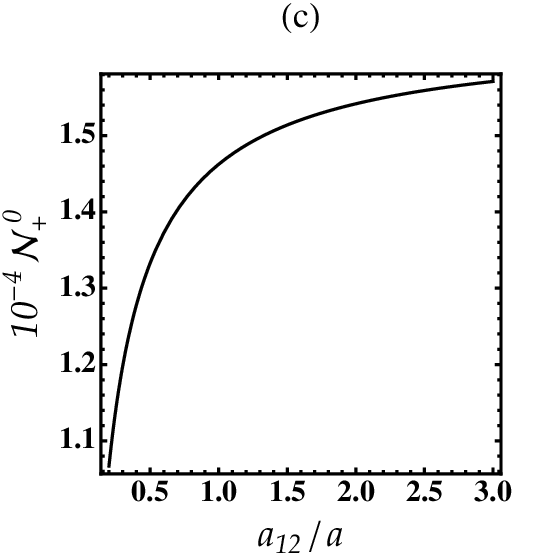}
\includegraphics[scale=0.45] {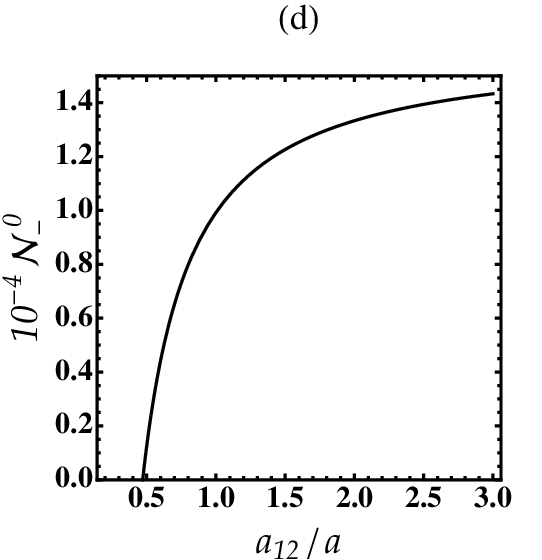}
 \caption{Decoherence factor associated with (a) the upper branch, $\Gamma_+^0$, and (b) the lower branch, $\Gamma_-^0$, 
for different values of the interspecies interaction, $a_{12}/a$.
Non-Markovianity  measure associated with (c) the upper branch, ${\cal N}_+^0$,  and (d) the lower branch, ${\cal N}_-^0$, as a function of $a_{12}/a$.
Parameters are: $p=0.5$, $\alpha=0.76$, and $L=0.75$. }
 \label{SQb}
\end{figure}

\begin{figure}
\includegraphics[scale=0.46] {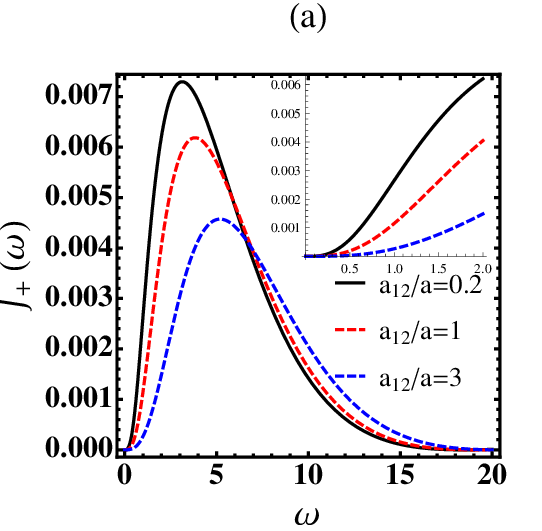}
\includegraphics[scale=0.46] {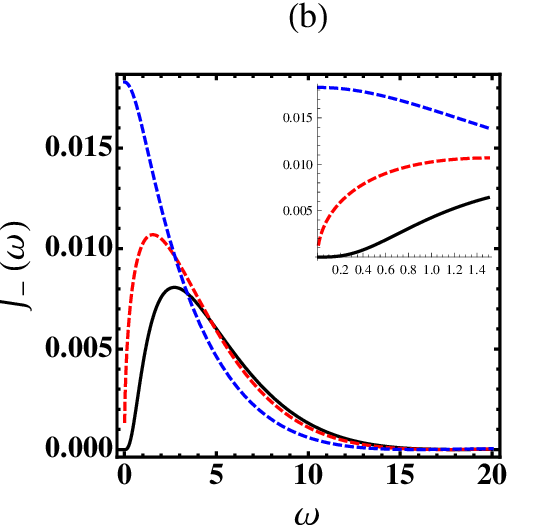}
\includegraphics[scale=0.46] {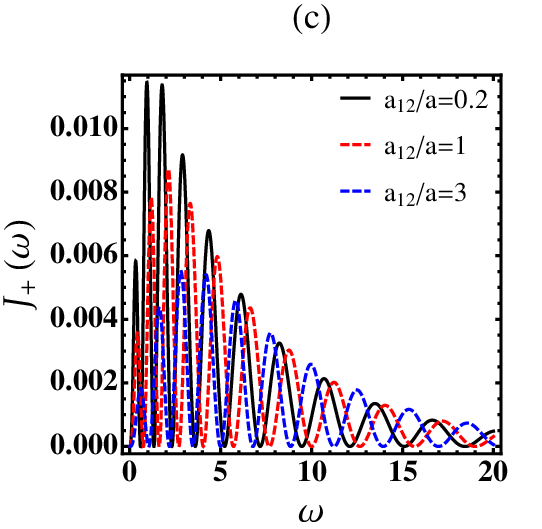}
\includegraphics[scale=0.46] {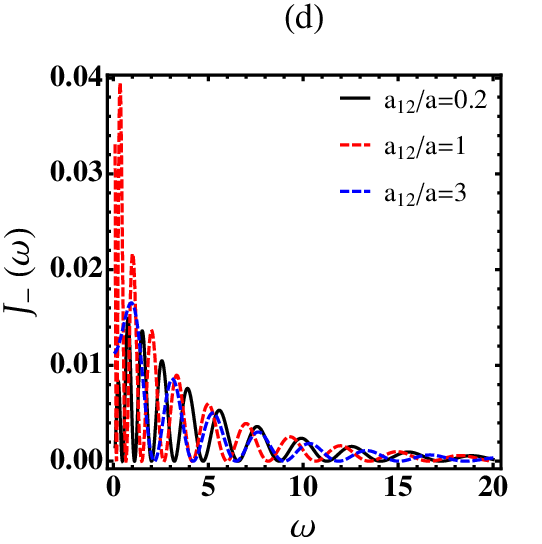}
 \caption{The spectral density functions (a) $J_+\omega)$  and (b) $J_-(\omega)$ for different values of  $a_{12}/a$ with $L=0.75$.
(c) and (d) The same as (a) and (b) but for $L=7.5$.
Parameters are the same as in Fig.~\ref{SQb}. The inset shows $J_{\pm}\omega)$ for very small frequencies.}
 \label{SF}
\end{figure}

In Fig.~\ref{SQb} we plot the decoherence factor and non-Markovianity  measure of both branches using the above experimental parameters and 
working in the limit of a continuum of modes, $\sum_k \rightarrow  (\ell /\pi) \int  dk$.
We see that the qubit retains its coherence at long times due to the suppressed effect of soft modes. This behavior becomes more obvious when 
the reservoir consists of the upper branch of the mixture where $\Gamma_{+}^0$ saturates quickly  for small times, $t \gtrsim 4$,
while $\Gamma_-^0$ saturates at $t \gtrsim 6$ (see Fig.~\ref{SQb} (a) and (b)). 
Another important remark is that for weak interspecies interactions $a_{12}/a =0.2$, the decoherence factor  $\Gamma_-^0$
is 2 times larger than $\Gamma_{+}^0$ and it is almost monotonic in time implying that the lower branch reservoir can give only some coherence back to the qubit.

Let us now move to the analysis of the non-Markovianity measure, which is defined to be the maximal amount of information that the system may recover from its environment, 
${\cal N}= \text{max}_{\rho_{1,2} (0)} \int_{\sigma >0} dt \sigma [t; \rho_{1,2} (0)]$.
Indeed calculating ${\cal N}$ is a complicated task due to the difficult optimization over all pairs of initial states.
However, for the purely dephasing impurity-Bose-Bose mixture that is considered in this work, the trace distance between these two states takes the form 
${\cal D}_{\pm}(t)= e^{-\Gamma_{\pm} (t)}$. 
This immediately gives the following expression of the non-Markovianity measure for a dephasing qubit \cite{Haik1}
\begin{equation} \label{NMarc}
{\cal N}_{\pm}= \int_{\Gamma_{\pm}' (t) <0}  \Gamma_{\pm}' (s) ds.
\end{equation}
This equation shows the impacts of the background parameters such as the interspecies interactions on the dynamics of information flow.

The numerical results in Eq.~(\ref{NMarc}) are shown in Fig.~\ref{SQb}.
We observe that the non-Markovianity is increasing with the interspecies interactions in both branches. 
It is clearly visible from Fig.~\ref{SQb} (c)  that ${\cal N}_+ >{\cal N}_-$ revealing that the upper branch reservoir can supply information to the qubit and hence
amplify the fraction of recovered information flow.
Remarkably, for weak interactions, $a_{12}/a \lesssim 0.5$, the dynamics is fully Markovian in the lower branch ${\cal N}_-$ (see  Fig.~\ref{SQb} (d)) indicating
that the flow of information is always from the qubit to the reservoir.

To gain deeper understanding of the crossover between Markovian and non-Markovian processes we analyze the reservoir spectral density function $J(\omega)$ \cite{Haik} in the 
low-frequency regime.
In the latter, the spectral density exhibits a power-law behavior $J(\omega) \propto \omega^s$, where $s$ is the Ohmicity parameter \cite{Haik, Haik1}.
We recall that the spectrum is sub-Ohmic when $s < 1$, Ohmic if $s = 1$ and super-Ohmic for $s > 1$.
The spectral density function characterizing the dephasing dynamics of a qubit in an ultracold Bose mixture environment
is defined as: $J_{\pm}(\omega)= \sum_k |g_{\pm k}|^2 \delta (\omega - \omega_{\pm k})$. 

For a symmetric Bose mixture, the behavior of $J_{\pm}(\omega)$ is displayed in Fig.~\ref{SF}. 
We see from Fig.~\ref{SF} (a) that as $a_{12}/a$ increases, the peak of $J_+(\omega)$ gradually shifts downward, giving rise to an increase in 
the value of $s$. Hence the spectrum changes from Ohmic for relatively small $a_{12}/a$ to super-Ohmic for $a_{12}/a \gtrsim 3$ in the regime of small frequencies 
(see the inset of Fig.~\ref{SF} (a)).
In such a case the environment can feed information back to the qubit and the dynamics is almost non-Markovian as foreseen above.

The situation is quite different for $J_-(\omega)$.
We observe from Fig.~\ref{SF} (b) that as $a_{12}/a$ increases, the peak of $J_-(\omega)$ is shifting upward indicating that the
spectrum changes from sub-Ohmic for $a_{12}/a=0.2$ to super-Ohmic for $a_{12}/a=1$  (see the inset of Fig.~\ref{SF} (b)).
Remarkably,  $J_-(\omega)$ displays a distinct behavior for larger interspecies interactions, $a_{12}/a=3$ allowing 
a transition from the non-Markovian to a Markovian dynamics as shown in Fig.~\ref{SQb} (d).

Another very interesting phenomenon captured in Figs.~\ref{SF} (c) and \ref{SF} (d) is that the spectral density functions, $J_{\pm}(\omega)$, 
oscillate with $\omega$ for large separation between the two wells, $L=7.5$. 
Large amplitude oscillations are observed in $J_+(\omega)$ for strong interspecies interactions and small $\omega$. The situation is inverted for $J_-(\omega)$.
These oscillations are damped for large enough frequencies, $\omega$ regardless of the interspecies interaction strength.

Similar to the model comprising of a trapped impurity qubit immersed in a single BEC reservoir \cite{Haik1}, the spectral density functions $J_{\pm}(\omega)$ of both components vanish for
some specific values of $\omega$, signifying that some modes of the Bose mixture reservoir are completely decoupled
from the qubit which is indeed natural since the two wells diverge from each other.

In the phonon regime (low frequencies), one has to substitute the spectrum (\ref{sound}) into the sensor-reservoir coupling parameter $g_{\pm k}$ and obtain the Ohmic reservoir spectrum
density function:
\begin{equation} \label{SDF}
J_{\pm} (\omega)= \eta \,\omega \sin \left(\sqrt{2} \gamma L \frac{\omega}{\omega_{c \pm}} \right) e^{ -\omega^2/\omega_{c \pm}^2}, 
\end{equation}
where $ \omega_{c \pm}= \sqrt{2} c_{s \pm}/l_I$ is the cutoff frequency of each branch, $L= L/l_0$, and
$\eta= \left[n a_{I}^2 l_0^3 (m_I+m)^2/( \sqrt{2} \pi (l_I^2+l_0^2)^2 m_I^2) \right]  (na)^{-3/2}$.

\section{Effects of collective decoherence of two qubits}\label{twoq}

\begin{figure}
\includegraphics[scale=0.47] {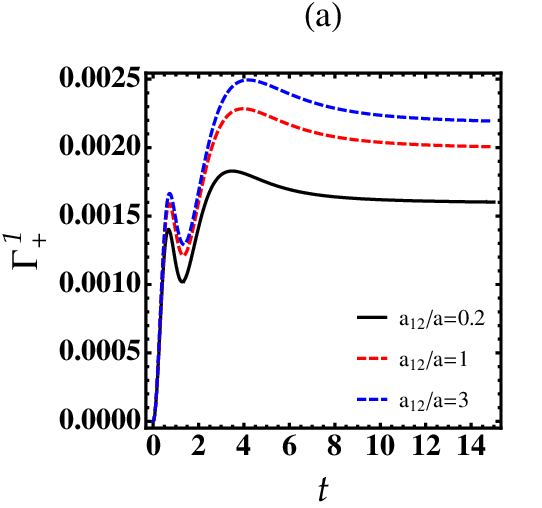}
\includegraphics[scale=0.45] {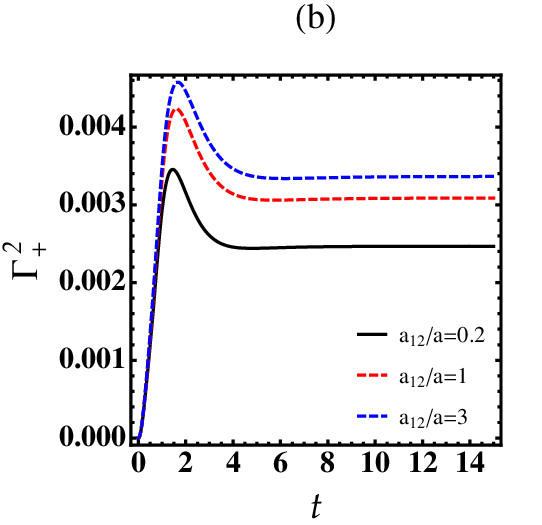}
\includegraphics[scale=0.47] {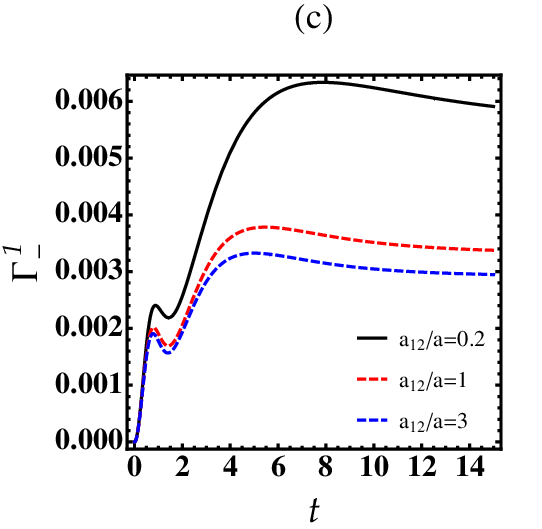}
\includegraphics[scale=0.45] {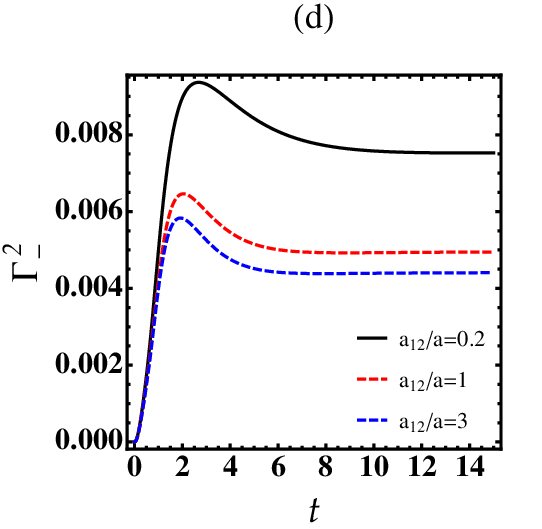}
 \caption{Decoherence factors (a) $\Gamma_{+}^1$, (b) $\Gamma_{+}^2$, (c) $\Gamma_{-}^1$, and (d) $\Gamma_{-}^2$ for different values of the interspecies interaction, $a_{12}/a$, 
with $2d=4L$.
Parameters are: $p=0.5$, $\alpha=0.76$, and $L=0.75$.}
 \label{CDs}
\end{figure}

\begin{figure*}
\includegraphics[scale=0.46] {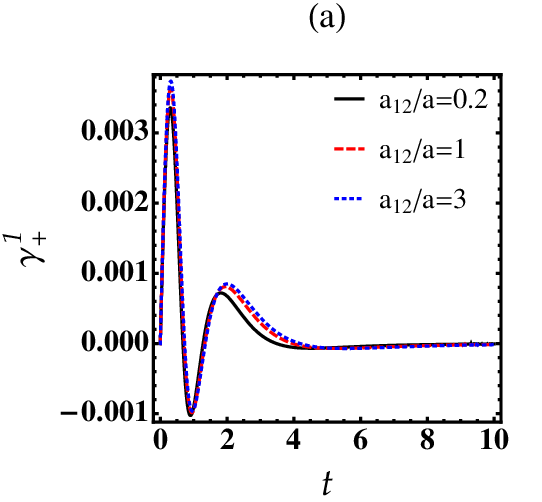}
\includegraphics[scale=0.45] {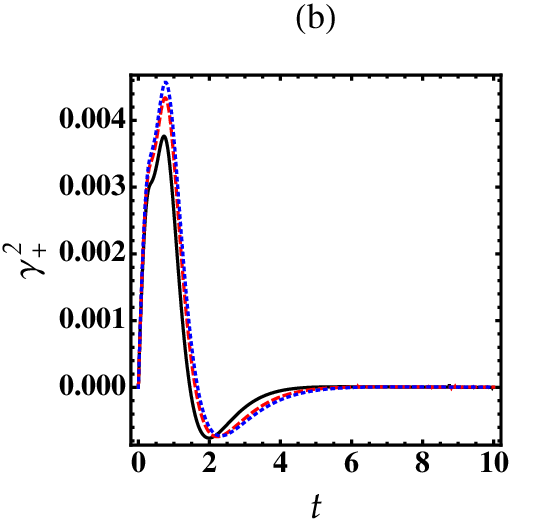}
\includegraphics[scale=0.45] {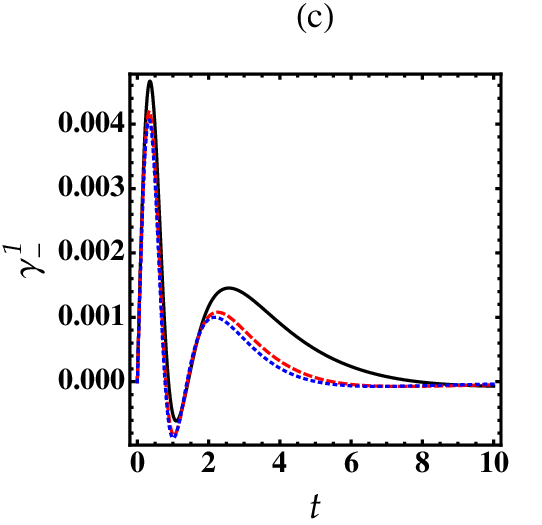}
\includegraphics[scale=0.45] {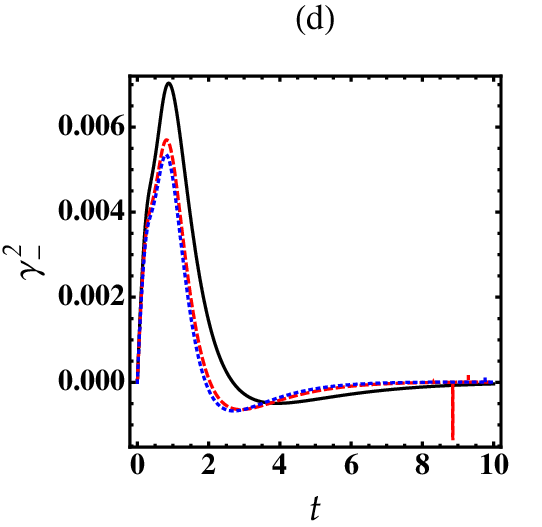}
\includegraphics[scale=0.46] {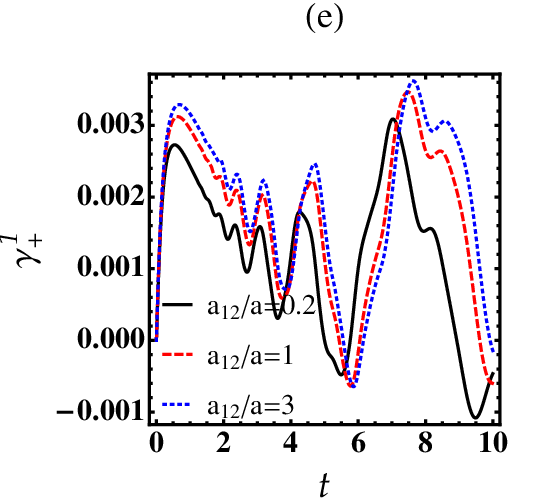}
\includegraphics[scale=0.46] {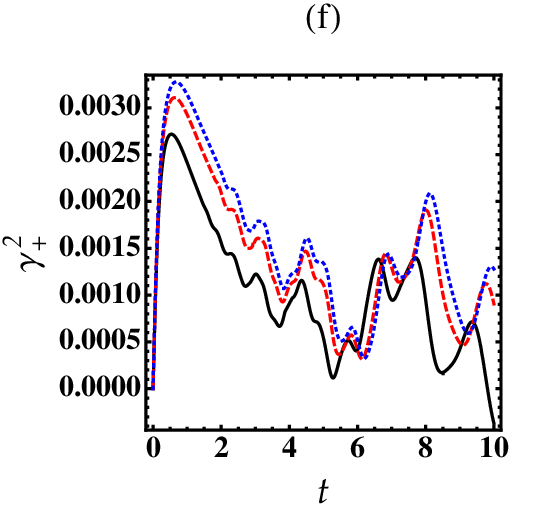}
\includegraphics[scale=0.45] {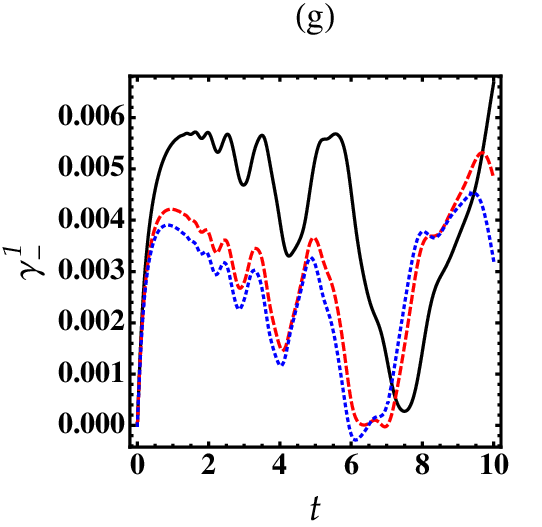}
\includegraphics[scale=0.45] {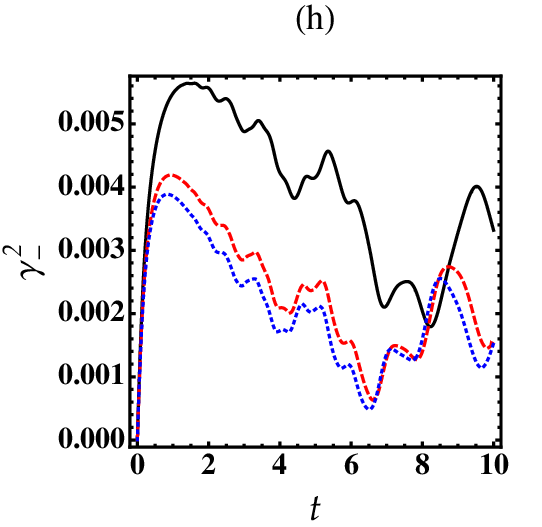}
 \caption{Time-dependent decay rates, $\gamma_{\pm}^{1,2} (t)$, for different values of the interspecies interaction, $a_{12}/a$ with $2d=4L$.
(a)-(d) $L=0.75$ and (e)-(h) $L=7.5$.
Parameters are the same as in Fig.~\ref{CDs}.}
 \label{DR}
\end{figure*}

Fascinating features manifest when one considers the collective decoherence of a system of two qubits (see Fig.~\ref {schm}). 
We look in particular at how the collective decoherence parameters, and the corresponding non-Markovianity and spectral density vary
in terms of interspecies interactions and the inter-qubit separation.

In such a case three decoherence parameters appear in the density matrix elements namely: $\Gamma_{\pm}^0$, $\Gamma_{\pm}^1$, and $\Gamma_{\pm}^{2}$ 
which are related to the elements such as: $|\rho_{I}^{0,0;0,1} (t)|= e^{-\Gamma^0(t)}\rho_{I}^{0,0; 0,1} (t) (0)$ or  $|\rho_{I}^{0,1; 1,1} (t)|= e^{-\Gamma^0(t)}\rho_{I}^{0,1; 1,1} (t) (0)$,
$|\rho_{I}^{0,0; 1,1} (t)|= e^{-\Gamma^1(t)}\rho_{I}^{0,0; 1,1} (t) (0)$, and $|\rho_{I}^{0,1; 1,0} (t)|= e^{-\Gamma^2(t)}\rho_{I}^{0,1; 1,0} (t) (0)$, respectively.
They correspond to the decoherences between states with the atoms on the same or on opposite sides of the double well \cite{Ciro}.
For two impurities at distance $2d = 4L$, the $\Gamma_{\pm}^1 (t)$ and $\Gamma_{\pm}^2 (t)$ read:
\begin{align} \label{DecFdw1}
\Gamma_{\pm}^1 (t)= \sum_{k} \frac{ 2g_{1\pm  k}^2} {\omega_{\pm k}^2} \sin^2 \left(\frac{ \omega_{\pm k} t}{2} \right) ,
\end{align}
where
\begin{equation} \label{coup}
 g_{1\pm k}= g_{I} \sqrt{\frac{n}{\ell}} \sqrt{\frac{E_{k}}{\hbar\omega_{\pm k}} } e^{-(k l_I/2)^2} \sin(k L) \cos(k d).
\end{equation}
and 
\begin{align} \label{DecFdw2}
\Gamma_{\pm}^2 (t)= \sum_{k} \frac{ 2g_{2\pm  k}^2} {\omega_{\pm k}^2} \sin^2 \left(\frac{ \omega_{\pm k} t}{2} \right),
\end{align}
where
\begin{equation} \label{coup}
 g_{2\pm k}= g_{I} \sqrt{\frac{n}{\ell}} \sqrt{\frac{E_{k}}{\hbar\omega_{\pm k}} } e^{-(k l_I/2)^2} \sin(k L) \sin(k d).
\end{equation}
We should stress that if the qubits are sufficiently close to each other, they will be coupled via correlation arising through an interaction mediated between them and the environment.
As the separation between the qubits increases, the correlation becomes weak and the qubit pairs are dissociated.
Therefore, they interact independently with their environments. In such a case the collective decoherence factors reduce to $\Gamma_{\pm}^{1,2} (t) \approx 2\Gamma_{\pm}^0 (t)$, 
where $\Gamma_{\pm}^0 (t)$ is the single qubit decoherence factor defined in Eq.~(\ref{DecFdw}).

Figure \ref{CDs} depicts that in a system of two qubits the decoherence factors  $\Gamma^1$ and $\Gamma^2$ of both branches saturate at a certain time to a stationary value 
which means that the qubits keep some of their coherence.
We see also that $\Gamma_{\pm}^1$ exhibit two peaks.  The positions and the heights of these peaks depend on the interspecies interactions strength.
The emergence of the second peak is associated with additional qubit coupling that modifies the behavior of the spectral density and  the Markovianity of the system.
Furthermore,  Fig.~\ref{CDs} shows the presence of sub- and superdecoherent states in  $\Gamma_{\pm}^1$ and $\Gamma_{\pm}^2$, respectively.
This can be attributed to the different contributions of the superpositions of the states $|0,0\rangle $, $|1,1\rangle $, $|0,1\rangle $ and $|1,0\rangle $ to
the decoherence factors \cite{Ciro}.
As in the case of a single qubit, $\Gamma_{-}^{1,2}$ are 2 times larger than $\Gamma_{+}^{1,2}$ indicating that the upper branch remains more robust against decoherence 
than the lower excitation component.

As we stated above, the evaluation of the non-Markovianity is often challenging even for a single-qubit system in a pure-dephasing scenario.
Therefore, the formula in Eq.~(\ref{NMarc}) is not suitable in practice for multi-qubit systems, as it requires searching for the optimal value across all states.
Signatures of non-Markovianity in the dynamics of our two-qubit system are contained in the time evolution of the decay rates \cite{Addis,Lu}, which are defined as:
\begin{equation} \label{2QDR}
\gamma_{\pm}^{1,2}= \frac{d\,\Gamma_{\pm}^{1,2}}{dt}.
\end{equation}
If $\gamma_{+}^{1,2}$ or  $\gamma_{-}^{1,2}$ are always positive, the dynamics is Markovian in both branches.
Conversely, if at least one of these time-dependent decay rates takes negative values, then the system displays non-Markovianity.

In Fig.~\ref{DR}, we plot the time-dependent coefficients $\gamma_{\pm}^{1,2}$ for different values of the interspecies interaction, $a_{12}/a$, and different inter-well distances.
For a small distance between the qubits, $L=0.75$, $\gamma_{\pm}^{1,2}$ take negative values in certain short intervals of time for any value of $a_{12}/a$ implying that the dynamics 
is non-Markovian in both branches (see Figs.~\ref{DR} (a)-(d)). 
It is clearly visible that $\gamma_{\pm}^{1,2}$ exhibit maxima and minima, then they saturate at zero, revealing that the dephasing of the qubits is halted. 
Note that this behavior holds true for a system consisting of two-qubit interacting with a single BEC reservoir \cite{Addis}.

Increasing the distance between wells, $L=7.5$, leads to a reduction in the quantum correlations between qubits. 
The functions $\gamma_{+}^{1,2}$ have negative values at long times giving rise to the occurrence of a non-Markovian dynamics in the upper branch (see Figs.~\ref{DR} (e) and \ref{DR} (f)).
Obviously,  the decay rates related to the lower branch, $\gamma_{-}^{1,2}$, are positive during their time evolution, 
thus the dynamics of the qubits is almost Markovian (see Figs.~\ref{DR} (g) and \ref{DR} (h)). This indicates  a continuous loss of information from the qubits to the lower branch environment.
As also shown in Fig.~\ref{DR} (g) (see blue the line), for relatively strong interactions, $a_{12}/a=3$, $\gamma_{-}^1$ is brought to negative values at $t\simeq 6$,
implying that the lower branch of the first component exhibits a non-Markovian behavior upon increasing interspecies interaction.

\begin{figure}
\includegraphics[scale=0.46] {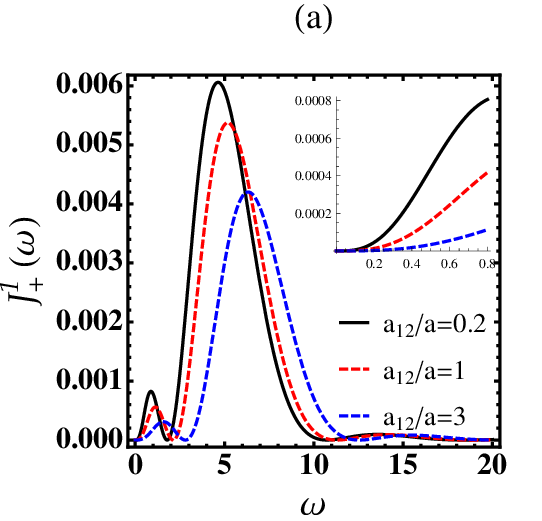}
\includegraphics[scale=0.46] {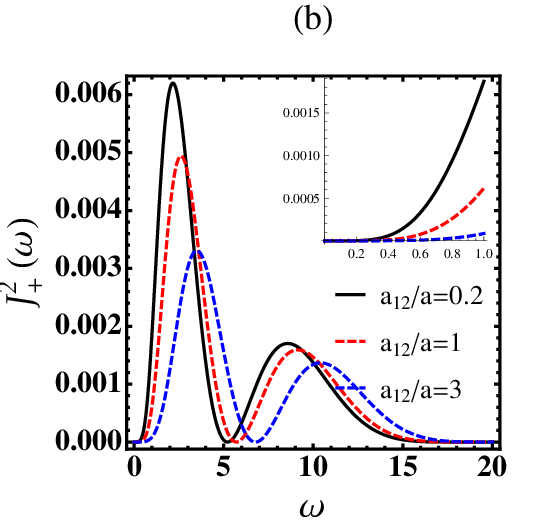}
\includegraphics[scale=0.46] {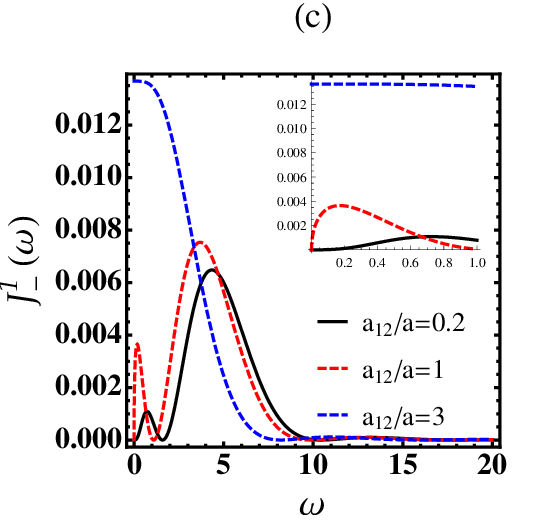}
\includegraphics[scale=0.46] {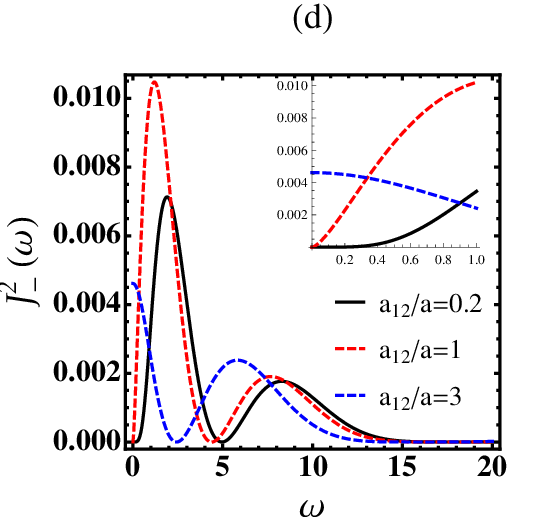}
 \caption{The spectral density functions (a) $J_{+}^1(\omega)$, (b) $J_{+}^2(\omega)$,  (c) $J_{-}^1(\omega)$, and (d) $J_{-}^2(\omega)$ for different values of $a_{12}/a$.
Parameters are the same as in Fig.~\ref{CDs}. The insets show $J_{\pm}^{1,2}(\omega)$ for very small frequencies.}
 \label{CSF}
\end{figure}

Figure \ref{CSF}  points out that oscillations appear in both $J_{\pm}^1(\omega)\equiv\sum_k |g_{1\pm}|^2 \delta(\omega-\omega_{\pm k})$ and 
$J_{\pm}^2(\omega)\equiv\sum_k |g_{2\pm}|^2 \delta(\omega-\omega_{\pm k})$ due to the peculiar collective character of the coupling.  
We also see that for small frequencies, $J_+^{1,2} (\omega)$ are almost super-Ohmic  for all the values of $a_{12}/a$ considered here (see Fig.~\ref{CSF} (a) and \ref{CSF}(b)) 
reflecting the creation of non-Markovian effects in the dynamics of the two qubits.
However, $J_-^{1} (\omega)$ is super-Ohmic for weak interspecies interaction ($a_{12}/a=0.2$) and sub-Ohmic for relatively intermediate intercomponent interactions ($a_{12}/a=1$).
The spectral density $J_-^{2} (\omega)$ is Ohmic for weak $a_{12}/a$ while it is super-Ohmic for $a_{12}/a=1$.
Increasing further the interspecies interaction ($a_{12}/a=3$), $J_-^{1,2} (\omega)$ display an anomalous behavior revealing a transition to Markovian dynamics.
It is clearly seen that few modes decouple from the qubits (i.e.  $J_{\pm}^1(\omega)$ and $J_{\pm}^2(\omega)$  cancel for certain critical values of $\omega$) 
even for a small separation between the two wells, which is in contrast to the single qubit case.

\begin{figure}
\includegraphics[scale=0.46] {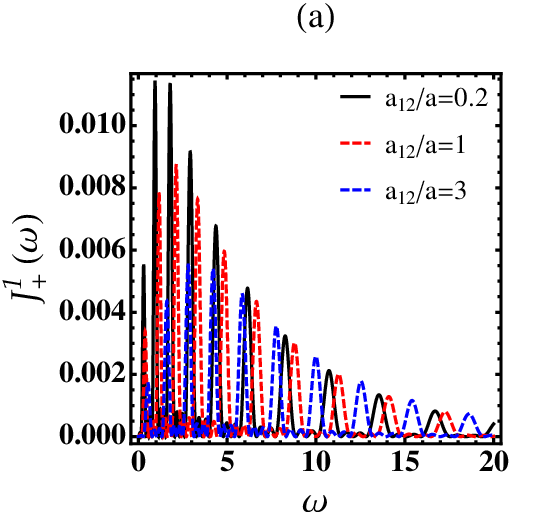}
\includegraphics[scale=0.46] {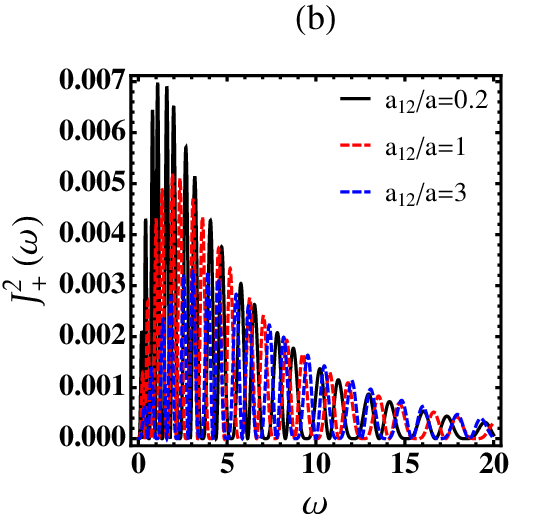}
\includegraphics[scale=0.46] {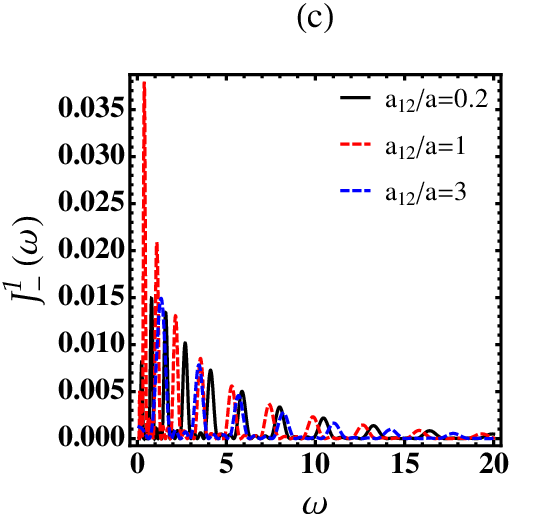}
\includegraphics[scale=0.46] {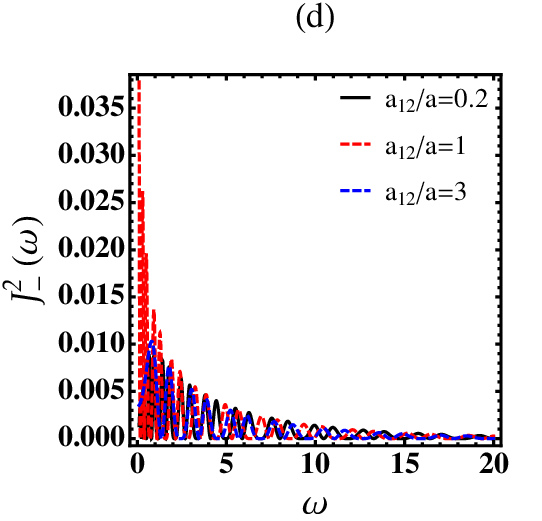}
 \caption{The same as in Fig.~\ref{CSF} but for $L=7.5$.}
 \label{LCSF}
\end{figure}

For completeness, let us also examine the role of the separation between wells in the behavior of the spectral density.
Changes in the separation between wells have a crucial effect on the spectral density in the whole range of frequencies where more pronounced oscillations appear 
in $J_{\pm}^{1,2}$ as shown in Figs.~\ref{LCSF}.
Regarding $J_+^{1,2}(\omega)$, they respond by oscillating with increasing amplitudes as $a_{12}/a$ gets decreased in the limit $\omega \gtrsim 10$
while they oscillate with the same amplitudes in the regime of large frequencies.
The situation is reversed for $J_-^{1,2}(\omega)$.
It is noteworthy that $J_{\pm}^1(\omega)$ oscillate periodically with small and large amplitudes according
to the interspecies interaction strength and to impurity-environment coupling. 
We observe on the other hand that in the limit of high frequencies, $J_{\pm}^1(\omega)$ and $J_{\pm}^2(\omega)$ oscillate, but their peak amplitude profile saturates to small values 
signaling a weak connection between qubits and their bath.

\section{Reservoir induced Entanglement} \label{Entg}

Based on Eq.~(\ref{EO}), the interaction between two qubits, mediated by the BECs environment, induces a coupling term of the form $U_{\pm}(t) \sim \mathcal{J}_{\pm}(t)\hat 
\sigma_z^1 \hat \sigma_z^2$, where ${\mathcal{J}}_{\pm}(t)$ represents the 
environment-induced coupling strength, given by:
\begin{equation}
{\mathcal{J}}_{\pm}(t)= 2\sum_k \frac{f_{k \pm}(t) g_{\pm 
k}^2\cos(2kd)}{\varepsilon_{\pm k}^2},
\end{equation}
with $g_{\pm k}$ as given in Eq.~(\ref{coup}). This interaction generates entanglement between the two qubits. To analyze this 
process, the system is initialized in the product state:
$
|\psi (0) \rangle = \frac{1}{2}(|0\rangle + |1 \rangle)_1 \otimes (|0\rangle + 
|1 \rangle)_2,
$
where each qubit is prepared in an equal superposition of its basis states. 
The entanglement dynamics between the qubits is quantified using the concurrence, a widely adopted entanglement measure for two-qubit systems.

The concurrence for each component, \( C_{\pm}(\rho_{\pm}) \), is derived from the density matrix operators of the two-qubit system, \( \hat \rho_{\pm}(t) \), 
defined as:
$\hat \rho_{\pm}(t) = \text{Tr}_B \left[\hat U(t)^\dagger \left(  |\psi(0\rangle \langle \psi(0)|\otimes \rho^B_{\pm} \right) \hat U(t)\right].$
By substituting the thermal equilibrium state, $\rho^B_{\pm}$, of the environment, 
the explicit form of the density matrix operator in the basis $\{|0,0 \rangle, |0,1\rangle, |1,0 \rangle, |1,1 \rangle \}$ is obtained as follows :
\begin{equation} \label{DensM}
	\hat \rho_{\pm}(t) = \frac{1}{4}
	\begin{pmatrix}
		1 & r({{\mathcal{J}}_{\pm}}) & r({{\mathcal{J}}_{\pm}}) & 
		e^{-\Gamma_{\pm}^{1}} \\
		r^*({{\mathcal{J}}_{\pm}}) & 1 & e^{-\Gamma_{\pm}^{2}} & 
		r^*({{\mathcal{J}}_{\pm}}) \\
		r^*({{\mathcal{J}}_{\pm}}) & e^{-\Gamma_{\pm}^{2}} & 1 & 
		r^*({{\mathcal{J}}_{\pm}}) \\
		e^{-\Gamma_{\pm}^{1}} & r({{\mathcal{J}}_{\pm}}) & 
		r({{\mathcal{J}}_{\pm}}) & 1
	\end{pmatrix},
\end{equation}
where \( r({\mathcal{J}}_{\pm}) = e^{2i \mathcal{J}_{\pm}} e^{-\Gamma_{\pm}^{0}} \). Here, we have neglected some trivial phase factors, 
as they do not contribute to the generation of entanglement. 
From Eq.~(\ref{DensM}), we can clearly see that the BEC environment plays two distinct roles in the dynamical evolution of the qubits. 
First, it induces an interaction between the qubits in a unitary manner, generating entanglement. 
Second, through dissipation, it degrades the entanglement between the qubits~\cite{Tan3, Tan4}.

To calculate the concurrence, we  apply the spin-flip operation to obtain the matrix 
$\hat{\tilde\rho}_{\pm}= \hat\sigma_y \otimes \hat\sigma_y \hat\rho^*_{\pm} 
\hat\sigma_y \otimes \hat\sigma_y$, where $\hat\sigma_y$ is the Pauli-$y$ 
matrix, and $\hat\rho^*_{\pm}$ is the complex conjugate of $\hat\rho_{\pm}$. 
The eigenvalues $\lambda_{\pm}^{(n)}$ of the matrix ${\hat{\tilde 
\rho}_{\pm} \hat\rho_{\pm}}$ are then computed and sorted in decreasing order.
The concurrence is subsequently obtained using the expression:
\begin{equation} \label{Con}
	C_{\pm}(\rho_{\pm})= \text{max} \left \{0,  \sqrt{\lambda_{\pm}^{(1)}} - 
	\sqrt{\lambda_{\pm}^{(2)}} - \sqrt{\lambda_{\pm}^{(3)}} - 
	\sqrt{\lambda_{\pm}^{(4)}} \right \},
\end{equation}
where $0 \leq C_{\pm} \leq 1$. This formula ensures non-negative concurrence, as required for a valid entanglement measure. When $C_{\pm} > 0$, the qubits are 
entangled, with larger values indicating stronger entanglement. Conversely, $C_{\pm} = 0$ indicates a separable (non-entangled) state.

\begin{figure}
\includegraphics[scale=0.34] {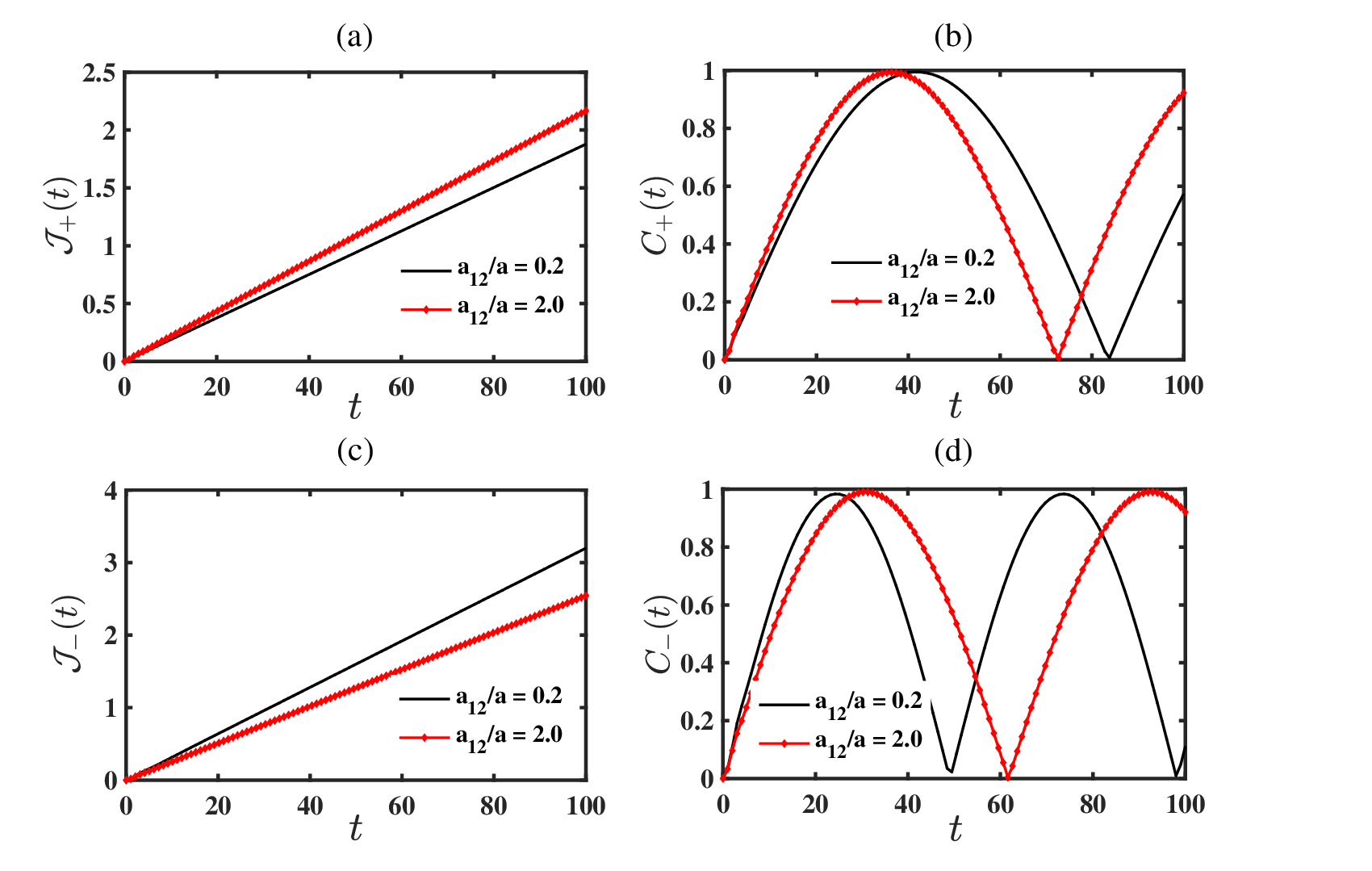}
\caption{ Time evolution of the environment-induced coupling strength, ${\mathcal{J}}_{\pm}(t)$, and  concurrence, $C_{\pm}(t)$, for different values of  $a_{12}/a$.
Here, $L=7.5$ and $2d=4L$.}
 \label{concur}
\end{figure}

The numerical results for the dynamics of the environment-induced coupling strength ${\mathcal{J}}_{\pm}(t)$ and the concurrence for various interspecies 
scattering lengths are presented in Fig.~\ref{concur}. 
As shown in Fig.~\ref{concur}(a) and \ref{concur}(c), the environment induces strong interactions between the qubits in both the upper and lower branches. 
For the upper branch, an increase in the value of $a_{12}/a$ enhances the coupling strength ${\mathcal{J}}_{\pm}(t)$, whereas the opposite trend is observed for 
the lower branch. In both cases, the interaction strength significantly exceeds the environment-induced dissipation effects, as demonstrated in Fig.~\ref{DR}, 
indicating the dominance of the unitary component of the environment's quantum effects.

Figures~\ref{concur}(b) and \ref{concur}(d) illustrate the entanglement dynamics as a function of the interspecies scattering length. These plots show that the 
environment-induced interaction increases the concurrence from 0, approaching 1 (maximum entanglement) with periodic oscillations. 
This suggests that the dissipation effects of the environment are nearly negligible. 
A comparison of Figs.~\ref{concur}(b) and \ref{concur}(d) reveals that for the upper branch, the period of $C_+(t)$ shortens as $a_{12}/a$ increases, leading to a quicker attainment of 
maximum entanglement, while the opposite trend occurs for $C_-(t)$ in the lower branch.

\section{Conclusions} \label{Conc}

In conclusion, we examined the dynamics of quantum information flow in one and two qubits trapped in a double-well potential immersed in a quasi-1D
ultracold Bose-Bose mixture reservoir. These systems generate interesting phenomena such as oscillations of coherence at finite times and the survival of coherence 
at long times, leading to a significant improvement in the accuracy of quantum dynamics.
Our numerical outcomes revealed that the system retains some coherence at long times in both branches.
We also explored the link between the form of the spectral density functions and the emerging qubit dynamics. 
The crossover from Markovian to non-Markovian dynamics which is observed in the lower branch is controlled only in the regime of low frequencies.
We demonstrated that the interspecies interactions and the distance between two wells play a pivotal role in the Ohmicity character and
thus in the non-Markovianity measure of each component.
Moreover, we found that for certain initially separable states, the correlations induced by two-component BECs can generate entanglement even between two largely distant qubits.
We showed that we can control the entanglement dynamics by adjusting system reservoir parameters such as the interspecies interaction in both components.
From a fundamental point of view, our results are crucial for quantum information processing.

\section*{Acknowledgments}
A.B.  acknowledges support from the Hassiba Benbouali University of Chlef. Q.S.T acknowledges support from the National Natural Science Foundation of China (NSFC) (Grant No. 12275077).

\end{document}